\documentclass{aastex63}

\accepted{April 29, 2020}

\submitjournal{AJ}

\shorttitle{Prospects for Directly Imaging Transiting Exoplanets}
\shortauthors{Stark et al.}

\begin{document}

\def\apj{ApJ}
\def\nat{Nature}
\def\aaps{A\&AS}
\def\pasp{PASP}
\def\aap{A\&A}
\def\procspie{Proc. of SPIE}
\def\apjl{ApJ}
\def\nat{Nature}
\def\aj{AJ}
\def\apjs{ApJS}
\def\mnras{MNRAS}

\title{Toward Complete Characterization: Prospects for Directly Imaging Transiting Exoplanets}

\correspondingauthor{Christopher C. Stark}
\email{cstark@stsci.edu}

\author{Christopher C. Stark}
\affiliation{Space Telescope Science Institute \\
3700 San Martin Dr. \\
Baltimore, MD 21218, USA}

\author{Courtney Dressing}
\affiliation{University of California, Berkeley, CA 94720, USA}

\author{Shannon Dulz}
\affiliation{University of Notre Dame, Notre Dame, IN 46556, USA}

\author{Eric Lopez}
\affiliation{NASA Goddard Space Flight Center, Greenbelt, MD 20771, USA}

\author{Mark S. Marley}
\affiliation{NASA Ames Research Center, Moffet Field, CA 94035, USA}

\author{Peter Plavchan}
\affiliation{George Mason University, Fairfax, VA 22030, USA}

\author{Johannes Sahlmann}
\affiliation{Space Telescope Science Institute, Baltimore, MD 21218, USA}

\begin{abstract}

High contrast direct imaging of exoplanets can provide many important observables, including measurements of the orbit, spectra that probe the lower layers of the atmosphere, and phase variations of the planet, but cannot directly measure planet radius or mass. Our future understanding of directly imaged exoplanets will therefore rely on extrapolated models of planetary atmospheres and bulk composition, which need robust calibration. We estimate the population of extrasolar planets that could serve as calibrators for these models. Critically, this population of ``standard planets" must be accessible to both direct imaging and the transit method, allowing for radius measurement. We show that the search volume of a direct imaging mission eventually overcomes the transit probability falloff with semi-major axis, so that as long as cold planets are not exceedingly rare, the population of transiting planets and directly imageable planets overlaps. Using current extrapolations of \emph{Kepler} occurrence rates, we estimate that $\sim$8 standard planets could be characterized shortward of 800 nm with an ambitious future direct imaging mission like LUVOIR-A and several dozen could be detected at V band. We show the design space that would expand the sample size and discuss the extent to which ground- and space-based surveys could detect this small but crucial population of planets.

\end{abstract}


\section{Introduction}
\label{intro}

Decades of advancement in the detection of extrasolar planets has produced a new understanding of exoplanet demographics. We now know that nature produces a vast array of extrasolar planets, from hot Jupiter-sized exoplanets on retrograde orbits to abundant sub-Neptunes for which there is no solar system analog. To understand the true nature of these planets we must compare observations with models. However, modeling the structure, composition, and atmospheric properties of extrasolar planets is complex and fraught with degeneracies---e.g., without a useful mass constraint, quantifying O$_2$ abundance and interpreting biosignatures on a directly-imaged potentially Earth-like planet could be difficult. 

Ideally we would be able to observe all of the planet's salient features, including mass, radius, orbit, and spectrum.  However, the types of measurements we can make are limited by the observation method.  The radial velocity method can place a lower bound on the planet's mass, and measure the orbital period and eccentricity, but cannot inform the planet's radius or atmospheric properties.  The transit method can measure inclination and orbital period, and can derive planet radius from a measurement of transit depth, but atmospheric spectra are often limited to the hottest planets by high-altitude hazes. Together, the radial velocity and transit methods can provide estimates of an exoplanet's mass, radius, orbit, and bulk density---critical inputs to models.

To observe the atmospheres of smaller, more temperate planets, and to probe deeper into the atmospheres of planets with hazes, we must move to direct imaging \citep{NAP25187}.  Direct imaging can provide spectra from the UV to NIR, revealing absorption features from many key species, including O$_2$, O$_3$, H$_2$O, CH$_4$, and CO$_2$.  Direct imaging can also directly measure the exoplanet's orbit, phase variations as the planet orbits the star, color variations in the planet's surface as it spins on its axis, and glint from specular reflection off of surface water, potentially revealing continents and oceans.  By obtaining a spectrum with broad wavelength coverage, direct imaging can place loose constraints on the planet's radius, but cannot directly measure either the radius or mass \citep[e.g.,][]{nayak2017}. 

While ground- and/or space-based radial velocity and astrometric measurements can in principle provide useful mass measurements for many planets over a wide variety of orientations, only those planets in edge-on orbits will allow for radius measurements via the transit method. This geometric requirement suggests that the number of exoplanets that are both transiting and accessible to future direct imaging missions is small. In this paper we estimate the number of exoplanets that meet these criteria using simple constraints on planet brightness and orbital properties. We show that while the expected number of transiting exoplanets that can be spectrally characterized to reasonably long wavelengths ($<$800 nm) via direct imaging is small, it is likely non-zero, and many may be available for photometric study at shorter wavelengths. These exoplanets would act as important calibrators for models, as all of their salient features could in principle be directly measured: mass, radius, orbit, atmospheric composition, and potentially surface properties.  We show the design space that would expand the sample size, and discuss the degree to which current and future surveys using ground- and space-based assets with a variety of methods could detect this small but crucial population of ``standard planets."

\section{The overlap of direct-imaging and transit detection methods}

Conventional wisdom suggests that the transit and direct-imaging methods probe separate exoplanet populations.  To first order, the probability of an exoplanet transit is given by $R_{\star}/a$, where $R_{\star}$ is the radius of the host star and $a$ is the semi-major axis of the planet, assuming circular orbits. As a result, exoplanets on large orbits transit rarely; an Earth twin has roughly a $0.5\%$ probability of transit.  It is no surprise that transit surveys have detected thousands of short period planets, but relatively few at periods longer than a year.  Direct-imaging, on the other hand, can only detect exoplanets outside of the inner working angle (IWA) of the coronagraph or starshade (loosely speaking).  Typical IWAs for future mission concepts like HabEx and LUVOIR are $\sim50$ mas \citep{habex_final_report, luvoir_final_report}; for a Sun-like star at 10 pc, this excludes planets interior to 0.5 AU.

These two opposing facts do not, however, suggest zero overlap between the methods.  Because the IWA of a direct imaging mission sets the maximum distance at which we can observe an exoplanet with orbital semi-major axis $a$, the volume of stars with planets accessible to direct imaging is roughly $\propto a^3$ (we note that this approximate scaling relationship may change slightly depending on the instrument throughput near the IWA).  Thus, as long as the occurrence rates of planets are not exceedingly low at large $a$, and the planets do not become exceedingly faint, the number of available systems will eventually overcome the $R_{\star}/a$ transit probability.

To estimate the number of standard planets accessible to a future direct imaging mission, we adopted the nominal planet occurrence rates for FGK stars from \citet{dulz2020}, expressed in terms of mass and semi-major axis. \citet{dulz2020} calculated the occurrence rates for a 100$\times$100 grid of bins covering planet masses from 0.08 M$_{\Earth}$--15 M$_{\rm J}$  and orbtial periods from $5.4$ days to 113 yrs. These occurrence rates are based on the community-averaged ExoPAG SAG13 values of Kepler occurrence rates \citep{sag13_report, kopparapu2018}, but substitute radial velocity occurrence rates at larger masses. Beyond a period of $\sim1$ yr, the SAG13 occurrence rates are an extrapolation and predict a dynamically unstable number of exoplanets. \citet{dulz2020} correct this extrapolation and account for dynamical stability by maximally packing systems while imposing a mutual Hill sphere stability criterion for periods beyond $\sim1$ yr. This admittedly first-order correction ignores the possibility of additional packing of exoplanets in resonant chains. As a result, we consider the occurrence rates of \citet{dulz2020} a reasonable, and possibly conservative, estimate at long periods. 

We adopted the stellar target list of \citet{stark2019}, extended to 100 pc. To first order, this target list is the Hipparcos catalog vetted using Gaia distances.  Because the Hipparcos catalog is only complete to $V\sim8$, many M stars beyond $\sim15$ pc are missing. While this is of no concern for yield estimates focusing on direct imaging surveys for potentially Earth-like exoplanets, here it will underestimate the number of detectable large, cold exoplanets. We therefore consider our target list to be conservative.

For each star in the target list, we calculated the probability of a standard planet existing by multiplying the occurrence rates by the transit probability.  Transit probabilities were calculated using stellar radii derived from $B-V$ color, using Eq. 2 and Table 1 from \citet{boyajian2014}.  In the absence of a measured spectral type dependence of the Kepler occurrence rates, we could either assume the semi-major axis distribution of occurrence rates is independent of spectral type, or the semi-major axis distribution scales with the square root of the stellar luminosity.  The former biases $\eta_{\Earth}$ to be higher for earlier type stars, while the latter maintains an $\eta_{\Earth}$ independent of spectral type. We adopted the latter method to be consistent with recent exoplanet yield calculations \citep{stark2019}.

For each star, we applied four cuts to the transit probability map that express the performance of our future direct imaging mission.  First, we required all planets to be brighter than $\Delta {\rm mag}$ = 26.5 magnitudes fainter than the host star, motivated by the coronagraphic noise floors adopted by previous studies \citep{stark2019}. Here we note that because coronagraph performance should improve at larger separations, we consider this again a conservative estimate for the colder planets that will be of importance to this study.  

Second, we also required that the exoplanets be bright enough to have reasonable exposure times. Instead of directly calculating the exposure times of every planet for a specific mission concept, with an assumed spectral resolving power $R$ and SNR, we imposed a simple V band brightness cutoff.  We adopted the nominal assumption of $V<30$, which roughly translates to an $R=70$, SNR $=10$ spectrum for a larger mission like LUVOIR-A (later we vary this parameter to estimate the performance of other mission concepts and spectral data quality).  To estimate the $V$ band magnitude of each exoplanet, we used Equation 3 from \citet{brown2005} and assumed planets were at quadrature, had a Lambertian phase function, and geometric albedos of $0.2$ for $R\le1.4$ R$_{\Earth}$ and 0.5 for $R>1.4$ R$_{\Earth}$. Planet masses were converted to radii using the mass-radius relationship of \citet{chenkipping2017}.

Third, we required that the planets reside exterior to the IWA of a future direct imaging mission when at quadrature. The IWA can vary significantly based on the future mission architecture.  For our nominal calculations, we adopted an IWA = 40 mas, consistent with the LUVOIR-A Apodized Pupil Lyot Coronagraph (APLC) design evaluated at 800 nm. By ensuring that a spectrum can be obtained on all of these planets at wavelengths shorter than 800 nm, all planets can be probed for CH$_4$ at 730 nm \citep{lupu2016}.

Treating IWA as a hard threshold, which we do for all calculations in this paper, is appropriate for the LUVOIR-A APLC designs. However, we note that some other coronagraphs do not exhibit a hard IWA and instead feature a throughput that gradually rises from $\sim$1 $\lambda/D$ to the OWA \citep{pueyo2019}. For these coronagraphs, the minimum detectable planet brightness is a function of separation based on the instrument throughput. Our estimates can only be applied in an approximate sense to such instruments.

In addition to the above cuts for direct imaging, we required that the transits be deep enough and frequent enough for useful radius measurements. For our nominal calculations we adopted a noise level $N=30$ ppm over 1 hour, consistent with estimates for the \emph{PLATO} mission \citep{rauer2018}. For each exoplanet, we calculated the geometric transit depth $\delta$ and duration $t$ in hours, and required 
\begin{equation}
\label{transit_snr_equation}
 \delta > {\rm SNR_{\delta}}\, N\, (nt)^{0.5},
 \end{equation}
where $n$ is the number of transits estimated as $n=P'/P$, and $P'$ is the timeline (in years) over which future transit observations could be conducted to measure radius, nominally chosen to be 5 years; we require $P<P'$. We adopted a threshold SNR$_{\delta} = 5$, implying a $20\%$ radius precision based on photometric precision, a reasonable assumption given that most radius measurements will be systematically limited by the stellar radius measurement to $\sim10\%$.

\begin{figure}
\centering
\includegraphics[width=6in]{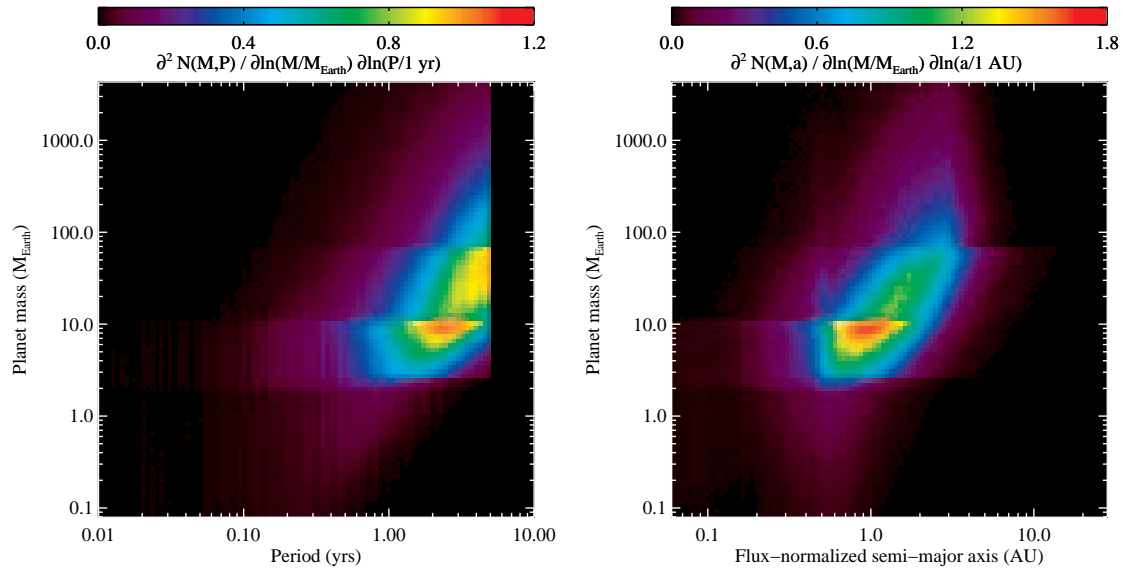}
\caption{Probability distribution of transiting exoplanets accessible to a future direct imaging mission with IWA=40 mas and limiting $V<30$ as a function of planet mass and period (left), and flux-normalized semi-major axis (right). A wide variety of planets are accessible, though probability is maximized at the limiting period of 5 years. \label{transit_prob_map_fig}}
\end{figure}

The left panel of Figure \ref{transit_prob_map_fig} shows the transit probability distribution summed over all $23000$ stars within 100 pc in our target list.  The discontinuity near $2.5$ $M_{\Earth}$ is due to the assumed change in geometric albedo, while other sharp vertical features are due to assumed changes in the mass-radius relationship, a discontinuity in the SAG13 occurrence rates at 3.4 $R_{\Earth}$, and the use of \citet{fernandes2019} occurrence rates for planets above $0.225$ Jupiter masses \citep{dulz2020}. Our adopted period cut at 5 yrs is clear as a horizontal discontinuity, as well as our photometric cut which removes planets to the lower-right of the plot.  As expected, the transit probability peaks near the maximum period in spite of the transit probability falling off as $a^{-1}$, as the number of accessible systems is maximized. The right panel shows the transit probability plotted as a function of mass and stellar flux-normalized distance; the transit probability distribution samples a wide range of planet masses and temperatures, but is maximized for Neptune-mass objects near the habitable zone (HZ).  Integrating over all masses and periods, we expect a total of $\sim$8 accessible standard planets for our nominal direct imaging assumptions. Roughly $1/3$ of these are $>30$ M$_{\Earth}$ and $1/3$ are $<8$ M$_{\Earth}$. 

Most of the standard planets shown in Figure \ref{transit_prob_map_fig} have periods in excess of 1 year, where the occurrence rates are poorly constrained. As a rough estimate on the uncertainty in the number of standard planets accessible with a mission like LUVOIR-A, we repeated our calculations using the pessimistic and optimistic occurrence rates of \citet{dulz2020} to find an expected value of $8^{+5}_{-3}$ accessible standard planets. We also repeated our calculations adopting the SAG13 nominal, pessimistic, and optimistic occurrence rates \citep{sag13_report} and arrived at a similar expected value of $7^{+8}_{-3}$ accessible standard planets.

\begin{figure}
\centering
\includegraphics[width=4in]{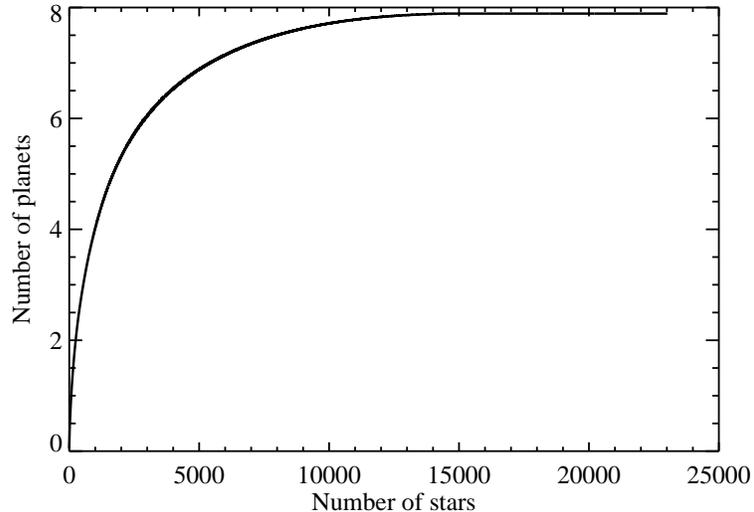}
\caption{Cumulative number of accessible standard planets from most to least productive stars. The majority of accessible planets occur around a few thousand stars. \label{cumulative_histo}}
\end{figure}

\begin{figure}
\centering
\includegraphics[width=6in]{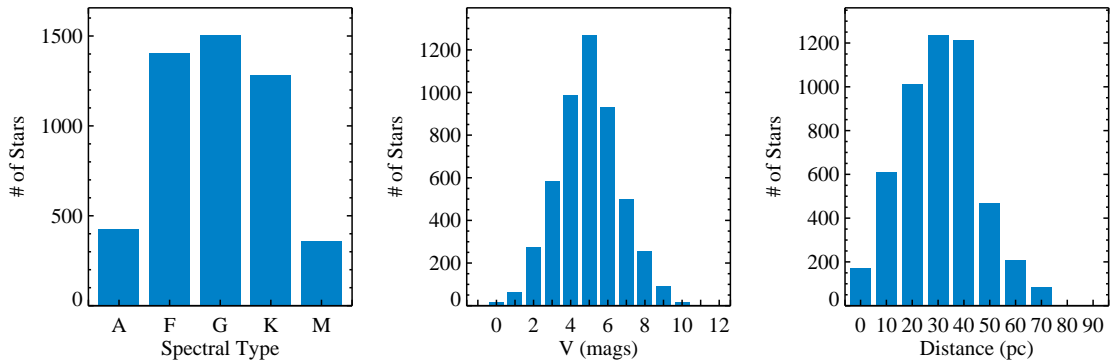}
\caption{Distribution of spectral types, apparent V band magnitudes, and distances for the first 5000 stars in Figure \ref{cumulative_histo}. \label{star_stats}}
\end{figure}

Not all stars contribute equally to Figure \ref{transit_prob_map_fig}. Figure \ref{cumulative_histo} shows the cumulative number of accessible standard planets for all stars, after sorting by each star's total probability. The majority of planets occur around just a few thousand stars. Figure \ref{star_stats} shows the spectral type distribution, V band magnitudes, and distances for the first 5000 stars. Most stars are FGK type and brighter than 7th magnitude, which may be in part due to the incompleteness of our input catalog \citep{stark2019}.

\begin{figure}
\centering
\includegraphics[width=4in]{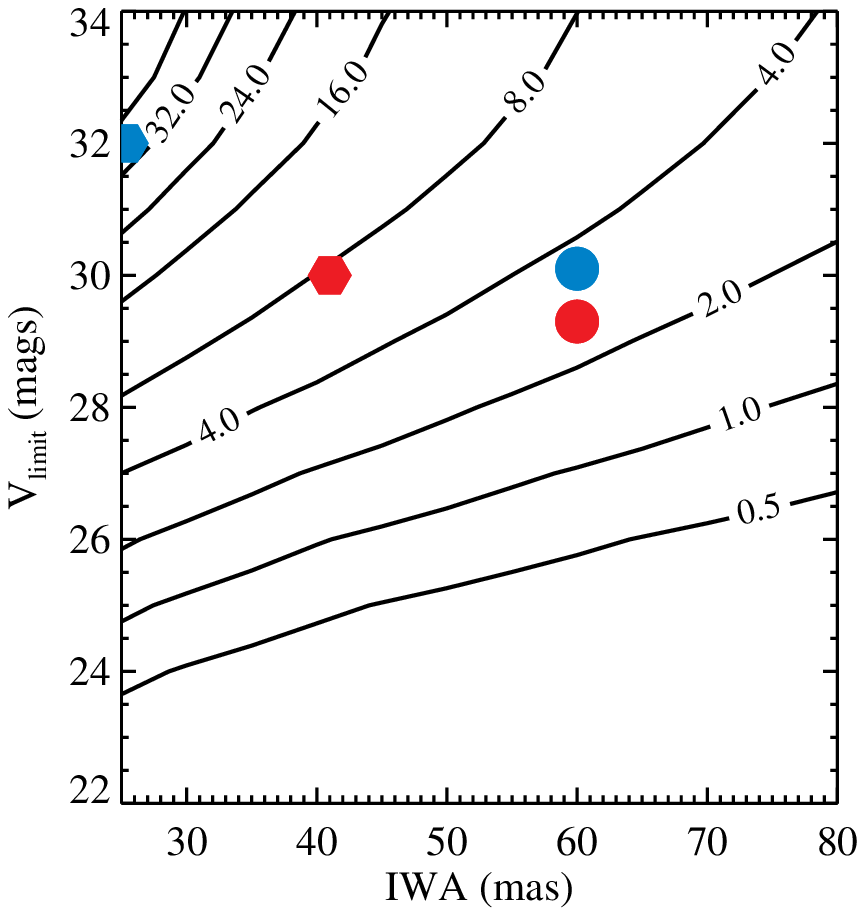}
\caption{Number of accessible standard planets as a function of the future direct imaging mission parameters. Spectral characterization and V band photometry with LUVOIR-A roughly correspond to the red and blue hexagons, respectively, while the red and blue circles roughly correspond to HabEx. \label{standard_sensitivity}}
\end{figure}

As one would expect, the capabilities of the future direct imaging mission have a significant impact on the number of accessible standard planets.  We repeated the above calculations for 12 values of IWA, ranging from 25 to 80 mas, and 12 values of limiting V band magnitude, ranging from 22 to 34. In all cases we kept the contrast noise floor constant at $\Delta {\rm mag}=26.5$. Figure \ref{standard_sensitivity} shows a contour plot of the total number of accessible standard planets as a function of these parameters.

Determining where any specific mission concept resides in Figure \ref{standard_sensitivity} requires some interpretation. E.g., if we require that all standard planets be characterizable with $R=70$, SNR$=10$ out to 800 nm, and set the integration cutoff time to 100 hours per bandpass including overheads, then LUVOIR-A is represented roughly by the red hexagon. If we require simple photometry with $R=5$, SNR$=10$ at 500 nm, then LUVOIR-A is represented by the blue hexagon, as the coronagraphic IWA scales with wavelength. HabEx, which has a wavelength-independent IWA of 60 mas from 300--1000 nm due to the use of a starshade, would be represented by the red and blue circles for each of these scenarios, respectively. Exposure times were estimated based on HabEx and LUVOIR-A design points using the study's exposure time calculators and a solar twin at 20 pc, 3 zodis of exozodi, and the planet on a 2 AU orbit at quadrature.

We note that the performance of the HabEx starshade, used for spectral characterization, is nearly ideal and operates near the diffraction limit of the telescope at 1000 nm. Therefore, we do not see a viable path to substantially move the red circle toward the upper-left without increasing the diameter of the telescope. The HabEx charge 6 vortex coronagraph, used for photometric detections, has half the throughput of the ideal coronagraph throughput \citep{belikov2019}, and the IWA could realistically be reduced by $\sim$10 mas, such that the blue circle could move closer to the contour of 8 planets.  LUVOIR-A, on the other hand, operates using an Apodized Pupil Lyot Corongraph with an on-axis telescope, which is farther from the ideal coronagraph performance \citep{belikov2019}. For LUVOIR-A, the coronagraph throughput could improve by a factor of $\sim$3, and the IWA could be reduced by $\sim$20 mas if coronagraph designs continue to mature for on-axis telescopes.  Both the red and blue hexagons could therefore conceivably reach double their current contours with future advancements in coronagraph design.

\section{Detecting standard planets}

Figure \ref{standard_sensitivity} suggests $\sim$8 standard planets could be \emph{accessible} for spectral characterization shortward of 800 nm to a future direct imaging mission like LUVOIR-A, and $\sim$3 dozen would be accessible at shorter wavelengths, but this does not mean they will be detected.  These low-probability transit events are scattered among thousands of stars. Below we discuss strategies for detecting these exoplanets.

\subsection{A byproduct of a direct imaging survey}
	
A future direct imaging mission like LUVOIR would likely conduct a large systematic survey for exoplanets via direct imaging. This survey would in large part be optimized for the detection and characterization of potentially Earth-like planets around stars within $\sim30$ pc. Along the way, such a survey would discover and measure the orbits of hundreds of other exoplanets.

To estimate how many standard planets would be detected by such a direct imaging survey, we adopted the nominal exoEarth-optimized survey from the LUVOIR Final Report \citep{luvoir_final_report}. This 2-yr survey observes $\sim$300 target stars with observation dates and exposure times chosen to maximize the yield of potentially Earth-like planets, not standard planets. Using these observations, we calculated the expected yields for exoplanets of all types, covering the full grid of occurrence rates from \citet{dulz2020}, following the methods of \citet{kopparapu2018, stark2019}. We then multiplied the expected direct imaging yields by the transit probability of each planet. We found that out of the $
\sim$600 expected directly imaged exoplanets from a 2-yr survey with LUVOIR-A, only $2$ are expected to transit, consistent with the cumulative yield of standard planets shown in Figure \ref{cumulative_histo} for a target list of $\sim$300 stars.

\subsection{A targeted survey} 

\subsubsection{Direct imaging method}

To increase the yield of standard planets from a direct imaging survey, one could optimize a direct imaging survey for those planets that are more likely to be standard planets.  Looking at the right panel of Figure \ref{transit_prob_map_fig}, we see that the probability distribution of accessible standard planets peaks in the warm Neptune region, and extends up to the warm and cold Jupiter-sized exoplanet regions. We ran three additional direct imaging yield calculations optimized for these planets using our target list expanded to 100 pc. Briefly, instead of distributing potentially Earth-like planets around every star, we distributed planets that fall within the warm Neptune, warm Jupiter, and cold Jupiter bins of \citet{kopparapu2018}, and optimized all aspects of the observations (selected stars, exposure times, number of observations, and delay time between observations) to maximize the yield of these planets using the methods of \citet{stark2019}. We then calculated the yield of all planet types when optimizing for any of these three sub-categories of planets. We limited the survey time to 6 months, including all overheads. We required detection of the planets, and did not budget for spectral characterization, but did require that $R=70$, SNR$=10$ spectral characterization times be $<100$ hours per bandpass. We also required at least 6 visits to every observed star to account for orbit determination. We find that additional 6 month surveys optimized for cold Jupiters, warm Jupiters, and warm Neptunes do not appreciably increase the number of standard planets. We conclude that a reasonable direct imaging survey is unable to detect more than a small handful of these standard planets.

\subsubsection{Transit method}

\begin{figure}
\centering
\includegraphics[width=6in]{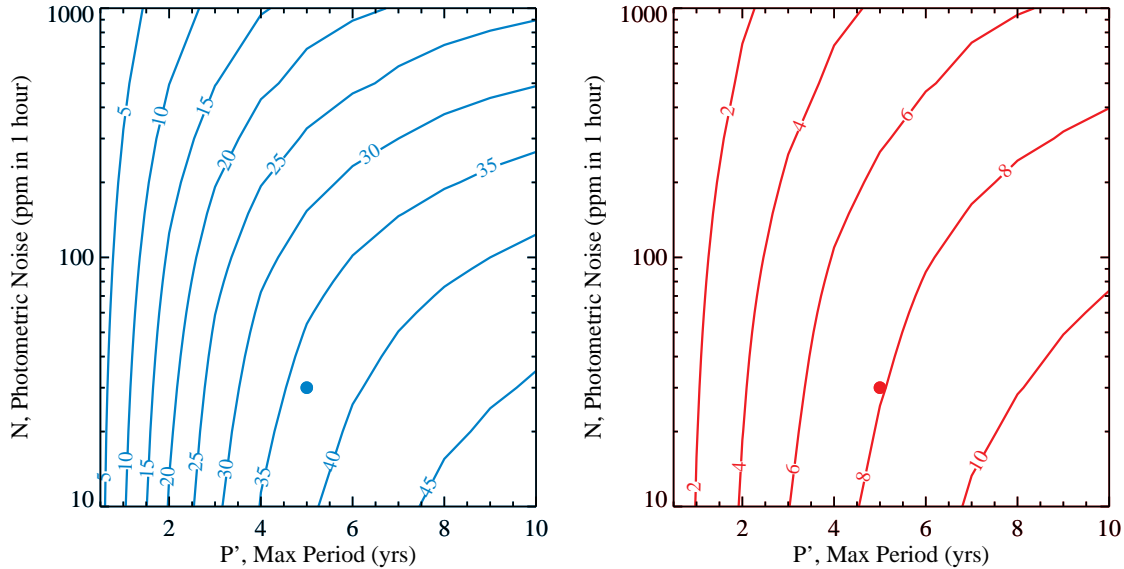}
\caption{The number of standard planets accessible to LUVOIR-A at $\lambda<500$ nm (left) and $\lambda<800$ nm (right) as a function of two primary transit survey parameters assuming full sky coverage; fractional sky coverage would reduce the numbers accordingly. \label{transit_survey_fig}}
\end{figure}

So far we have examined how the number of standard planets varies as a function of direct imaging capabilities for fixed transit detection capabilities ($N = 30$ ppm, $P'=5$ years). Here, we investigate how a transit survey could be used to detect standard planets while fixing the direct imaging capabilities, nominally chosen as representative of LUVOIR-A. We recalculated the number of standard planets accessible to such a mission as a function of $N$ and $P'$, as shown in Figure \ref{transit_survey_fig}. The left panel shows the number of standard planets that would be detectable at V band by LUVOIR-A while the right panel shows the number characterizable by LUVOIR-A to 800 nm. The dots indicate our nominal transit parameters.

Both panels of Figure \ref{transit_survey_fig} assume full sky coverage for the duration of the survey. The number of detected planets would be reduced approximately by the fraction of sky continually monitored during the transit survey. While the photometric noise of the transit survey could be relaxed to 100 ppm from our nominal 30 ppm without significant impact to the number of detected planets, full sky coverage is necessary. 

Nominally the TESS mission stares at a given patch of the sky for approximately one month and takes 2 years to survey nearly the full sky \citep{barclay2018}. The actual probability of a transit occurring in the TESS field of view is a complex result of overlapping sky coverage and period aliasing. However, roughly approximating it as the lesser of $(N_{\rm ext}+1)/(12P)$ and unity, where $N_{\rm ext}$ is the number of TESS 2-year extended missions and $P$ is the planet's period in years, we find that a total of 4 years of TESS surveys would find potentially 1 standard planet, mostly likely as a single transit event. Without a substantially extended mission and change of operations concept, the TESS mission will be unable to reliably find such planets.  PLATO will have the precision and cadence needed to find some of these planets, but certainly will not have full sky coverage for the duration of the mission \citep{rauer2018}. Only a long-duration, full-sky transit survey sensitive to long duration $\sim$100 ppm signals could recover a large fraction of standard plants. We conclude that the transit method will be unlikely to identify a significant number of standard planets.

\subsubsection{Radial velocity method}

To determine whether a radial velocity survey could detect an appreciable number of standard planets, we performed simple estimates of radial velocity surveys. Starting with the nominal LUVOIR-A mass-period histogram of standard planets for each star (the left panel of Figure \ref{transit_prob_map_fig} shows the sum over all stars), we calculated the radial velocity semi-amplitude $K$ of all planets in the histogram assuming circular orbits. We then determined the total number of planets with $K>K'$, where $K'$ is the limiting RV sensitivity. We assumed the radial velocity survey could detect all planets with orbital periods $P<5$ years assuming they met the RV sensitivity threshold. We considered only stars classified as spectral type G, K, and M, as the rapid rotation of stars above the Kraft break makes radial velocity measurements difficult. We then sorted stars by their total possible yield of standard planets with $K>K'$ to select the highest priority $N_{\star}$ stars.

\begin{figure}
\centering
\includegraphics[width=6in]{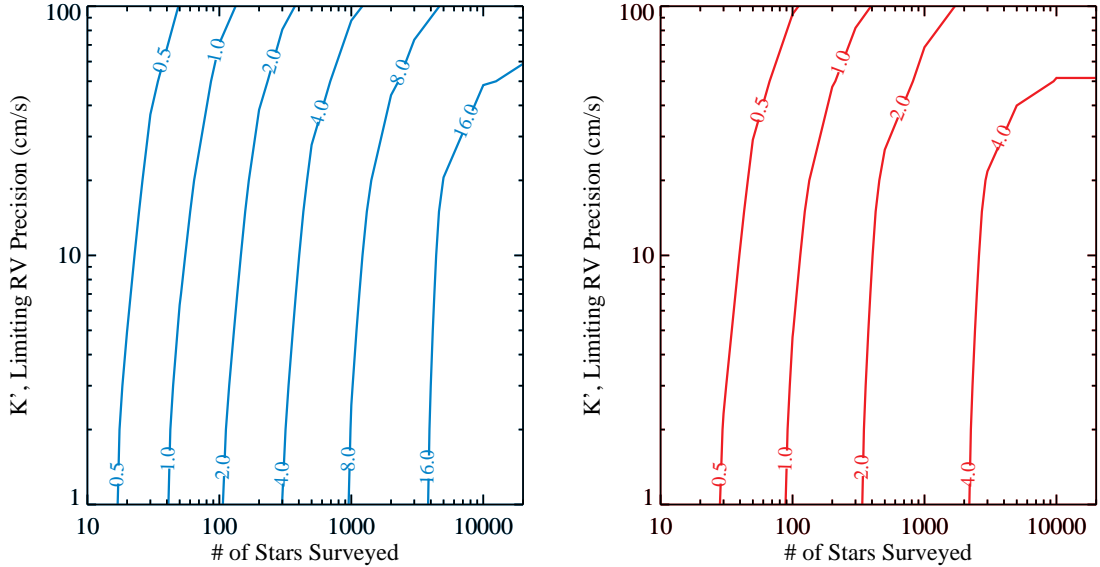}
\caption{The number of standard planets accessible to LUVOIR-A at $\lambda<500$ nm (left) and $\lambda<800$ nm (right) detected by a radial velocity survey with limiting RV semi-amplitude $K'$. \label{rv_survey_fig}}
\end{figure}

Figure \ref{rv_survey_fig} shows the expected yield of RV-detected standard planets as a function of $K'$ and $N_{\star}$. Because the standard planets shown in Figure \ref{transit_prob_map_fig} are predominantly $>10$ M$_{\Earth}$ and orbit at 1--3 AU, their radial velocities are typically in excess of 20 cm s$^-1$. The inability to survey F type and earlier stars limits the productivity of RV surveys to $\lesssim60\%$ of all possible standard planets. We note that ignoring the Kraft break to include all A and F type stars does not have a significant impact on the contour plots leftward of 1000 stars surveyed, but does increase the number of accessible standard planets when larger numbers of stars are surveyed.

An RV survey of a few hundred stars sensitive to $\sim40$ cm s$^-1$ signals could detect a handful of standard planets for V-band photometric study with LUVOIR-A. However, detecting standard planets amenable to $\lambda<800$ nm spectral characterization with LUVOIR-A would require an RV survey of thousands of stars to a precision of $\sim30$ cm s$^{-1}$. Seasonal aliasing would make the detection of planets with orbital periods near one year difficult. To reduce the impact of day/night aliasing, multiple facilities at multiple longitudes would be beneficial. Further, since RV measurements do not determine inclination, such a survey would identify many non-transiting planet candidates, necessitating follow-up transit measurements of a large number of stars to select candidates for direct imaging. We conclude that the radial velocity method will be unlikely to identify standard planets.

\subsubsection{Astrometric method\label{astrometry_section}}

Finally, we estimated the ability of astrometric surveys to detect standard planets. Again starting with the nominal LUVOIR-A mass-period histogram of standard planets for each star, we calculated the angular semi-major axis $\alpha$ of the star's barycenter motion for all planets in the histogram assuming circular orbits. We then determined the total number of planets with $\alpha>\alpha'$, where $\alpha'$ is the limiting astrometric survey sensitivity. Because the astrometric precision of Gaia depends on stellar brightness and is uncertain for the brightest stars (V$<$5) \citep{lindegren2018}, we also included a bright limit on the V band magnitude of the star, V$_{\rm bright}$. We assumed full sky coverage and sensitivity to planets with orbital periods as long as 5 years. 

\begin{figure}
\centering
\includegraphics[width=6in]{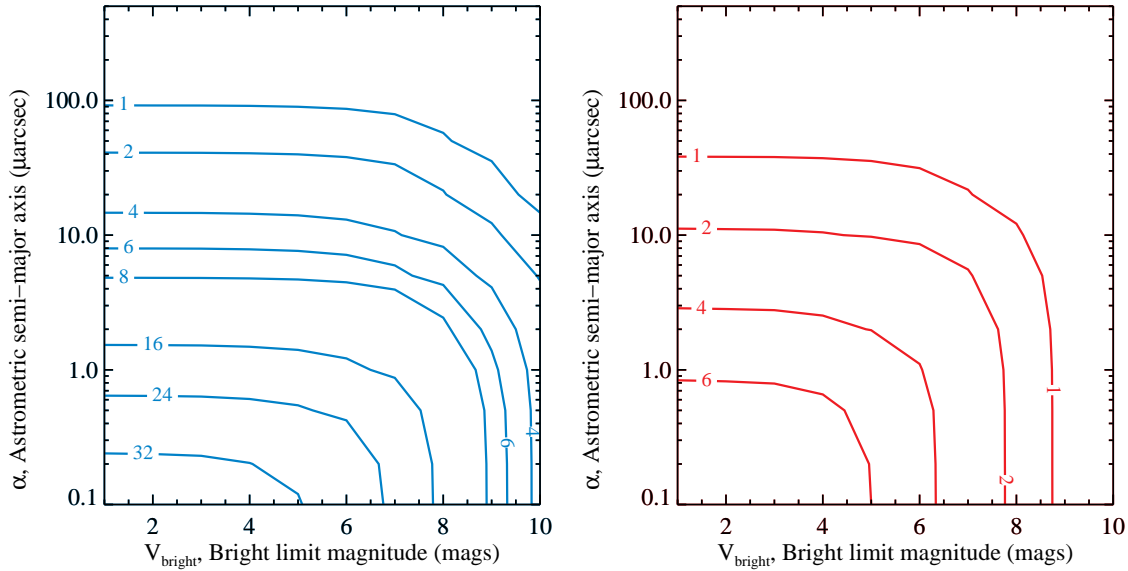}
\caption{The number of standard planets accessible to LUVOIR-A at $\lambda<500$ nm (left) and $\lambda<800$ nm (right) detected by an astrometric survey with limiting angular semi-major axis $\alpha'$ and bright limit V$_{\rm bright}$. \label{astrometric_survey_fig}}
\end{figure}

Figure \ref{astrometric_survey_fig} shows the expected yield of standard planets detected via an astrometric survey as a function of $\alpha'$ and V$_{\rm bright}$. Because standard planets are dominated by Neptune-mass planets at 1--3 AU around stars at tens of pc, typical astrometric signals are only $\sim$1 $\mu$as. Detecting these signal amplitudes is beyond the expected capabilities of Gaia \citep{perryman2014}. Assuming that the current formal precision limit of $\sim$50 $\mu$as at V$=5$ \citep{lindegren2018} can be matched in accuracy in the final Gaia data release, a detection threshold of S/N$>20$ \citep{sahlmann2015} translates into detectable signatures of $\alpha=120$ (80) $\mu$as for a 5 (10) year Gaia mission. The latter is still a factor of three larger than the predicted parameter space of standard planets accessible to LUVOIR-A at $\lambda< 800$ nm. An accuracy of $\sim$3--6 $\mu$as per Gaia measurement at V$=3$ would be required to reach into the parameter space of standard planets.

Unlike a potential RV survey, an astrometric survey would constrain the inclination of the planet. While certainly not precise enough to predict a transit, astrometric constraints of inclination should greatly reduce the number of candidate standard planets for future transit follow-up observations. We conclude that although Gaia is unlikely to identify standard planets, future all-sky astrometric surveys with higher accuracy are a promising avenue.

\subsubsection{Improving survey efficiency via inclination constraints}

We have shown that while the number of standard planets accessible with a future direct imaging mission may likely be non-zero, a large fraction of them are unlikely to be found using traditional survey approaches due to their rarity and moderately long periods. Here we speculate on an alternative approach.

First, we posit that if planets in a system tend to be coplanar, as has been observed for a number of compact multiplanet systems \citep[e.g.,][]{fang2012}, the number of systems to be searched could be greatly reduced if we had an estimate of their inclination. Under the assumption that planets and disks are coplanar to within an average dispersion $\sigma_i=10^{\circ}$ and we can measure the inclinations of systems perfectly, we could ignore 83\% of systems that are all more than 10$^{\circ}$ from edge-on. This reduces the target list from $\sim$5000 high priority stars to $\sim$800 stars. Such a reduced target list becomes more amenable to RV and ground-based transit surveys.

We consider four ways to estimate system inclination. First, direct imaging of a system's debris disk can provide the necessary precision on inclination. HST STIS routinely detects disks $\sim$1000$\times$ the density of the Kuiper Belt in $\sim$1000 seconds. A survey of 5000 stars to a depth of $1000\times$ the Kuiper Belt would require a full two months of exposure time. Of course only a small fraction of debris disks are that bright. In the limit of infinite contrast, detecting a disk as faint as the Kuiper Belt would take $1000\times$ as long, or roughly a week. A direct imaging survey of cold debris disks around a substantial fraction of the 5000 high priority stars is therefore implausible.

Second, we could rely on the detection of short period transiting planets, assuming that they are highly correlated with the inclination of the rest of the system. Specifically, we pose the following question: if a system has a standard planet, what are the odds that it also has a detectable short period transiting planet? Though we lack the proper statistical knowledge regarding system architectures and multiplicity to calculate this correctly, we can roughly estimate an answer by assuming only a correlation in inclination---i.e., the probability of a short period transiting planet is governed only by its mutual inclination to the standard planet. 

\begin{figure}
\centering
\includegraphics[width=6in]{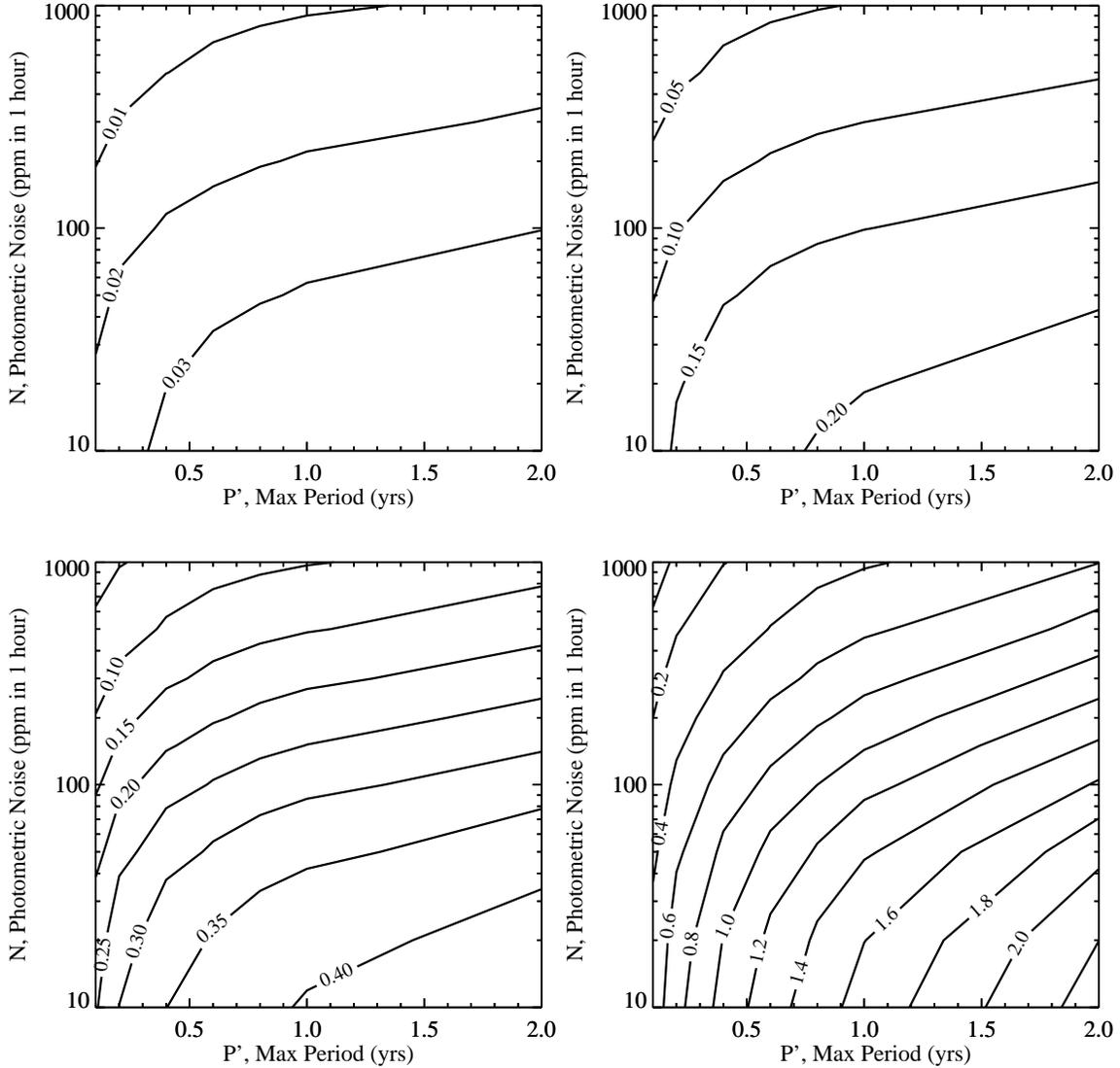}
\caption{Average number of detectable shorter period transiting planets per star for standard planet host stars assuming only inclination correlations. The four panels correspond to planet mutual inclinations uniformly distributed from 0-$90^{\circ}$ (top-left, no correlation), 0-$10^{\circ}$ (top-right), 0-$5^{\circ}$ (bottom-left), and $0^{\circ}$ (bottom-right, perfectly coplanar). For reasonable inclination dispersions, only a fraction of systems with standard planets will host detectable short period transiting planets. \label{short_period_transit_planet_completeness_fig}}
\end{figure}

We took the occurrence rate maps for the 5000 high priority stars from Figures \ref{cumulative_histo} and \ref{star_stats}, and calculated the number of detectable transiting planets per star under the assumption that every star had a standard planet and all systems have planets with orbital inclination dispersion $\sigma_i$. Assuming a uniform distribution of planetary inclinations from $-\sigma_i/2$ to $\sigma_i/2$ about the system midplane, the presence of a standard planet would mean that the rest of the planets in the system could have inclinations ranging from $-\sigma_i$ to $\sigma_i$ (allowing for the standard planet to be at the ``top" or ``bottom" of the inclination distribution). With this approximation, we recalculated the transit probability of every pixel in the occurrence rate map of each star by determining the number of orientations for which the planets transited. For given values of photometric noise $N$ and maximum survey period $P'$, we then added up the number of detectable transiting planets. Figure \ref{short_period_transit_planet_completeness_fig} shows the results: the average number of detectable transiting planets per star for systems that also host transiting standard planets as a function of $N$ and $P'$. The top-left plot of Figure \ref{short_period_transit_planet_completeness_fig} shows the average number of detectable transiting planets if there is no correlation of inclinations between planets; the odds of a shorter period transiting planet are very small. The top-right and bottom-left plots show the same for mutual inclinations uniformly distributed within 10$^{\circ}$ and $5^{\circ}$, respectively, and show that only a fraction of systems with standard planets will host detectable shorter period transiting planets. E.g., if planets are typically aligned to within $5^{\circ}$ of one another, the bottom-left plot suggests a full sky transit survey that achieves 100 ppm per hour on all planets with up to 6 month orbital periods would find transiting planets among roughly one quarter of all standard planet systems. Only when systems are perfectly coplanar (shown in the bottom right plot) is the probability of an inner transiting planet high. We note that these estimates ignore the possible effects of planet multiplicity. If multiplicity were included, the actual success rate may be higher.

\begin{figure}
\centering
\includegraphics[width=4in]{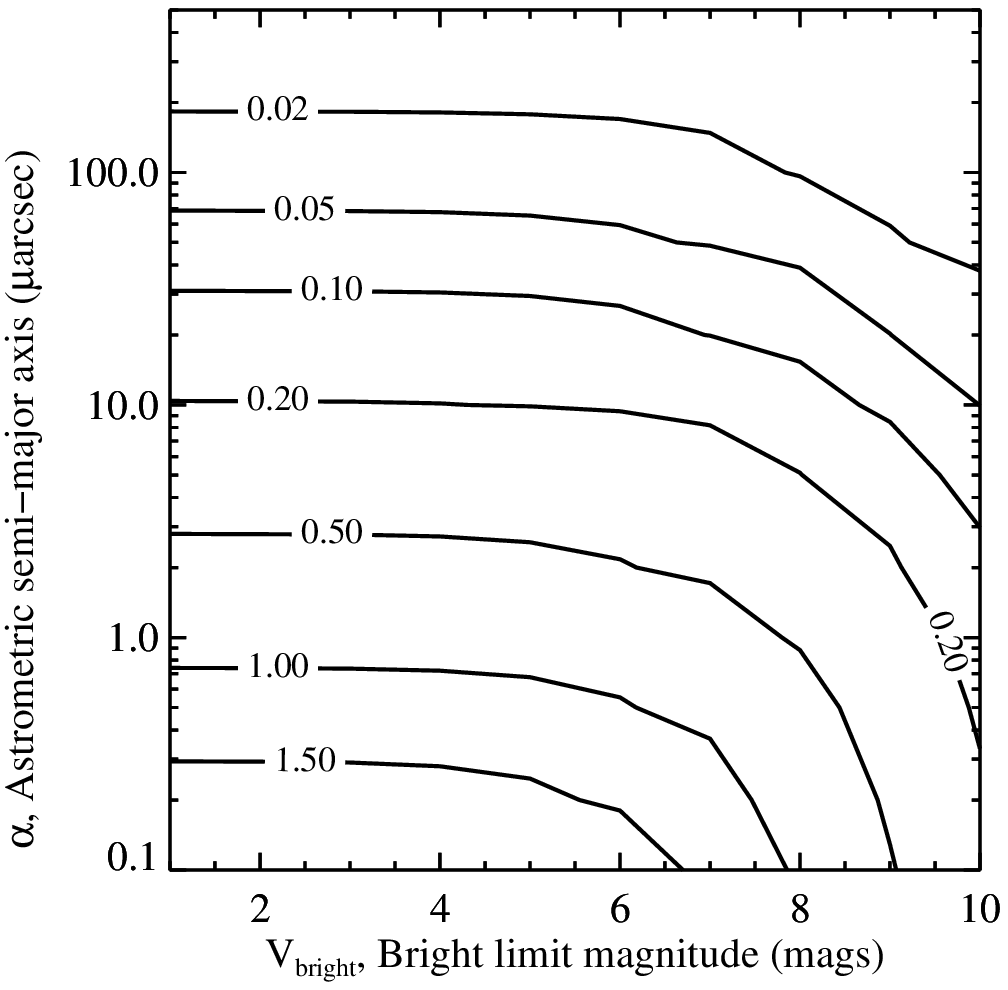}
\caption{Average number of astrometrically detectable planets per star for 5000 high priority systems that host a standard planet. Gaia would detect planets around a very small fraction of the high priority host stars.\label{astrometric_planet_completeness}}
\end{figure}

Finally, we consider using the astrometric method to estimate the inclinations of planetary systems. Specifically, we ask how many of the high priority 5000 stars will have planets detectable with Gaia. To answer this, we took the occurrence rate maps of every star, ignored the transit probability, and calculated the average number of planets per star detectable with Gaia assuming a maximum period of 5 years. For each pixel in the occurrence rate map, we calculated the astrometric signal as described in Section \ref{astrometry_section}. Figure \ref{astrometric_planet_completeness} shows the average number of planets detectable over the ensemble of 5000 stars as a function of astrometric survey parameters. Adopting $\alpha'=120$ $\mu$as and $V_{\rm bright}$=5 as representative Gaia performance, we see Gaia will detect planets around only a few percent of the 5000 high priority stars. We conclude that the inclination constraints provided by near term direct imaging, transit, and astrometric surveys are insufficient to inform standard planet target selection.

Measurement of stellar properties may also improve the efficiency of a search for standard planets. Measurement of stellar rotation axes could significantly aid such a search if they are highly aligned with planetary orbital axes. It is currently unclear whether this is typically the case \citep{winn2015}---observations show that some systems appear aligned \citep[e.g.,][]{wright2011,bergfors2011}, especially compact multiplanet systems \citep{campante2016}, while others appear to be significantly misaligned, such as the relatively rare population of hot Jupiters \citep[e.g.,][]{albrecht2012}. Co-alignment of planets with binary systems could also be helpful, but here the picture is equally unclear. The few circumbinary planets detected thus far appear to be aligned, though this may be a selection effect \citep{winn2015}. Further, ALMA observations show misaligned protoplanetary disks \citep{jensen2014} and theoretical studies suggest mechanisms to initiate and maintain $90^{\circ}$ misalignment \citep[e.g.,][]{farago2010,martin2017}.

\section{Conclusions}
\label{conclusions}

Future observations of directly imaged planets will produce detailed exoplanet spectra for a wide variety of planets, but our interpretation of those spectra will be dependent on models of planetary atmospheres. ``Standard planets" that are detectable with a wide variety of methods, enabling measurement of their radii, mass, and atmospheric spectra among other traits, could help calibrate these models. We showed that $\sim$8 standard planets are expected to be characterizable with LUVOIR-A shortward of 800 nm, while several dozen are detectable at V band. However, these planets are rare, scattered among thousands of stars, and largely beyond our current limits of detection. Only $\sim25\%$ of these planets are expected to be discovered during a direct imaging survey. None of the existing or upcoming space- or ground-based exoplanet surveys to date could appreciably increase that fraction, including TESS, Plato, Gaia, and upcoming RV surveys. To find a larger fraction of these planets, RV surveys would have to search thousands of stars for years to a level of $\sim$30 cm/s, transit surveys would have to cover the full sky with $\sim$100 ppm/hour precision for more than a decade, or full-sky astrometric surveys would have to achieve $\mu$as precision on the brightest stars for $\sim$5 years.

\acknowledgments

The authors would like to thank Daniel Fabrykcy and the anonymous referee for fruitful discussions and feedback. This work was partly supported by NASA's NExSS Virtual Planetary Laboratory funded by the NASA Astrobiology Program under grant 80NSSC18K0829. SDD acknowledges support by the National Science Foundation Graduate Research Fellowship Program under Grant No. DGE-1841556.

\bibliography{ms_v2.bbl}
\bibliographystyle{aasjournal}

\end{document}